\documentclass[12pt, twoside]{article}
\usepackage[utf8]{inputenc}
\usepackage[english]{babel}
\usepackage{amsthm, amssymb, amsmath}
\usepackage[utf8]{inputenc}
\usepackage{graphicx}
\usepackage{geometry}
\usepackage{indentfirst}
\usepackage{booktabs}
\usepackage{setspace}
\usepackage{array}
\usepackage[boxed]{algorithm2e}
\usepackage[table]{xcolor}
\usepackage{caption}
\usepackage{hyperref}
\usepackage{afterpage}
\usepackage{mathtools}
\usepackage{tikz}
\usepackage{nicefrac}
\usepackage{enumerate}
\usepackage{centernot}
\usetikzlibrary{arrows}
\graphicspath{ {./images/} }
\usepackage{relsize}
\usepackage[mathscr]{euscript}
\usepackage{float}
\usepackage{dblfloatfix}
\usepackage{url} 
\usepackage{cleveref}
\usepackage{subcaption}
\usepackage{cite} % add [super] to switch
\usepackage{lineno}  % For line numbering
\usepackage{tabularx}
\usepackage{lipsum}

\newcommand\blfootnote[1]{%
  \begingroup
  \renewcommand\thefootnote{}\footnote{#1}%
  \addtocounter{footnote}{-1}%
  \endgroup
}

\usepackage{titlesec}
\titleformat{\section}[block]{\bfseries\MakeUppercase}{\thesection}{1em}{\fontsize{14}{16}\selectfont}
\titleformat{\subsection}[block]{\bfseries\fontsize{12}{14}\selectfont}{\thesubsection}{1em}{}
\titleformat{\subsubsection}[block]{\itshape\fontsize{12}{14}\selectfont}{\thesubsubsection}{1em}{}

\definecolor{lightgrey}{rgb}{0.83, 0.83, 0.83}
\begin{document}
% \linenumbers
\begin{centering}
{\Large{Investigating the importance of county-level characteristics in opioid-related mortality across the United States}}\\[.5cm]

Andrew J. Deas\textsuperscript{1,2,*},
Adam Spannaus\textsuperscript{2},
Dakotah D. Maguire\textsuperscript{2},
% Heidi A. Hanson\textsuperscript{3}, 
Jodie Trafton\textsuperscript{3}, 
Anuj J. Kapadia\textsuperscript{2},
and Vasileios Maroulas\textsuperscript{1}\\[0.5cm]
\end{centering}
\begin{flushleft}
\textbf{Author Affiliations:}\\
\textsuperscript{1}Department of Mathematics, University of Tennessee, Circle Dr, Knoxville, 37916, TN, USA\\[0.25cm]

\textsuperscript{2}Computational Sciences and Engineering Division, Oak Ridge National Laboratory, Bethel Valley Road, Oak Ridge, 37830, TN, USA\\[0.25cm]

\textsuperscript{3}Office of Mental Health and Suicide Prevention, Veterans Health Administration, Willow Road, Palo Alto, 94025, CA, USA\\[0.5cm]

\textbf{*Corresponding author}:\\
\begin{center}
Andrew J. Deas\\
1 Bethel Valley Road, Oak Ridge, TN 37830\\
deasaj@ornl.gov\\[0.5cm]
\end{center}


\textbf{Highlights:}\\
$\bullet$ Introduces a novel imputation strategy for geospatial mortality data using queen-adjacent neighbors.\\
$\bullet$ Combines data analysis and machine learning to identify critical social factors in the opioid epidemic.\\
$\bullet$ Findings reveal structural vulnerabilities that may exacerbate opioid-related mortality when present at high levels, yet mitigate it when present at low levels.\\[0.5cm]

\textbf{Included in this pdf file:} main text, figures 1 through 8, tables 1 and 4

\blfootnote{Notice: This manuscript has been authored by UT-Battelle, LLC, under contract DE-AC05-00OR22725 with the US Department of Energy (DOE). The US government retains and the publisher, by accepting the article for publication, acknowledges that the US government retains a nonexclusive, paid-up, irrevocable, worldwide license to publish or reproduce the published form of this manuscript, or allow others to do so, for US government purposes. DOE will provide public access to these results of federally sponsored research in accordance with the DOE Public Access Plan (https://www.energy.gov/doe-public-access-plan).}

\newpage
\textbf{Abstract}\\[0.25cm]

The opioid crisis remains a critical public health challenge in the United States. Despite national efforts which reduced opioid prescribing rates by nearly 45\% between 2011 and 2021, opioid-related overdose deaths more than tripled during the same period. This alarming trend reflects a major shift in the crisis, with illegal opioids now driving the majority of overdose deaths instead of prescription opioids. Although much attention has been given to supply-side factors fueling this transition, the underlying structural conditions that perpetuate and exacerbate opioid misuse remain less understood. Moreover, the COVID-19 pandemic intensified the opioid crisis through widespread social isolation and record-high unemployment; consequently, understanding the underlying drivers of this epidemic has become even more crucial in recent years. To address this need, our study examines the correlation between opioid-related mortality and thirteen county-level characteristics related to population traits, economic stability, and infrastructure. Leveraging a nationwide county-level dataset spanning consecutive years from 2010 to 2022, this study integrates empirical insights from exploratory data analysis with feature importance metrics derived from machine learning models. Our findings highlight critical regional characteristics strongly correlated with opioid-related mortality, emphasizing their potential roles in worsening the epidemic when their levels are high and mitigating it when their levels are low.\\[0.25cm]

\textbf{Abstract word count:} 201\\[0.5cm]

\end{flushleft}

\newpage
\section{Introduction}

The opioid crisis in the United States remains one of the most daunting and complex public health challenges in recent history. Initially driven by prescription practices, national efforts and legislation to decrease the availability of opioids have significantly reduced prescriptions from 257.9 million in 2011 to 143.4 million in 2020, a 44.4\% decrease \cite{cdc_od_epidemic,ama_report,nih_report}. This decline in prescriptions has not led to a corresponding reduction in opioid-related overdose deaths however. On the contrary, opioid overdose death rates have surged within this same period \cite{pnas_mortality_forecasting}. In 2011, the rate stood at 7.3 per 100,000 persons, but by 2021, it had more than tripled to 24.7 per 100,000 persons \cite{death_rates_rising}.

These unsettling trends highlight a critical transformation in the opioid epidemic: Misuse is no longer primarily sustained through prescription medications but has shifted to illicit opioids such as fentanyl or heroin \cite{transition_to_illicit_opioids,fourth_wave}. This transition has been driven by several factors, including the increased availability of low-cost, high-potency synthetic opioids \cite{rise_of_synthetics_1,rise_of_synthetics_2}, the widespread mixing of drugs with unpredictable potency \cite{drug_mixing_1,drug_mixing_2,drug_mixing_canada}, the proliferation of online platforms enabling the distribution of illegal substances \cite{online_opioids_1,online_opioids_2}, and worsening regional conditions associated with preventable mortality linked to mental health and substance use\cite{deaths_of_despair_2015,deaths_of_despair_2017,deaths_of_despair_implications}. Although these factors had been exacerbating the epidemic well before 2020, the social isolation and record-high unemployment levels caused by the COVID-19 pandemic have further intensified the crisis \cite{pnas_covid19,covid_unemployment_loneliness,covid_unemployment_long_term_impact,social_isolation_covid_1,social_isolation_covid_2}.

While much attention has been given to the supply-side factors driving the opioid crisis, the underlying structural conditions that perpetuate and exacerbate opioid misuse are not as well understood. A growing body of research suggests that elements of the Social Vulnerability Index (SVI)---a composite measure designed to capture a community’s resilience to external stressors---play critical roles in shaping opioid-related outcomes \cite{svi_rates, opioids_and_svi_and_covid, opioids_and_svi_2}. Defined in \cref{table_svi_definitions}, the features in the SVI are county-level characteristics that have been identified as reliable predictors of opioid-related mortality \cite{svi_ml}; counties with elevated levels in these measures tend to exhibit higher rates of opioid use disorder (OUD) \cite{svi_oud}, and communities with worse housing conditions and less access to transportation often face reduced access to medications for OUD \cite{opioids_and_svi_2}.

\begin{table}[h!]
    \centering
    \begin{tabularx}{\textwidth}
    {>{\raggedright\arraybackslash}p{4cm} >{\raggedright\arraybackslash}p{4cm} >
    {\raggedright\arraybackslash}X}
        \toprule
        \textbf{Theme} & \textbf{Feature} & \textbf{Definition} \\
        \midrule
        \textbf{Socioeconomic Status} & Below poverty & Percentile percentage of persons below poverty estimate \\
        & Unemployment & Percentile percentage of civilian (age 16+)
        unemployed estimate \\
        & No high school diploma & Percentile percentage of persons with no high school diploma (age 25+) estimate  \\
        \midrule
        
        \textbf{Household Characteristics} & Age 65 and older & Percentile percentage of persons aged 65 and older estimate \\
        & Age 17 and younger & Percentile percentage of persons aged 17
        and younger estimate  \\
        & Single-parent households & Percentile percentage of single-parent households with children under 18 estimate \\
        & Limited English ability & Percentile percentage of persons (age 5+) who speak English ``less than well" estimate \\
        \midrule
        
        \textbf{Racial and Ethnic Minority Status} & Minority Status & Percentile percentage minority (Hispanic or Latino (of any race); Black and African American, Not Hispanic or Latino; American Indian and Alaska Native, Not Hispanic or Latino; Asian, Not Hispanic or Latino; Native Hawaiian and Other Pacific Islander, Not Hispanic or Latino; Two or More Races, Not Hispanic or Latino; Other Races, Not Hispanic or Latino) estimate \\
        \midrule
        
        \textbf{Housing Type and Transportation} & Multi-unit structures & Percentile percentage of housing structures with 10 or more units estimate  \\
        & Mobile homes & Percentile percentage mobile homes estimate \\
        & Crowding & Percentile percentage of households with more people than rooms estimate \\
        & No vehicle & Percentile percentage of households with no vehicle available estimate \\
        & Group quarters & Percentile percentage of persons in group quarters estimate (e.g., correctional institutions, nursing homes) \\
        \bottomrule
    \end{tabularx}
    \caption{Definitions of the county-level characteristics utilized in this study.}
    \label{table_svi_definitions}
\end{table}

While previous studies have provided important insights into the relationship between these county-level characteristics and opioid-related mortality, significant limitations persist. Many rely on cross-sectional data that does not capture trends over consecutive years and focus on aggregate measures rather than individual components. To address these gaps, this study utilizes a nationwide county-level dataset spanning consecutive years from 2010 to 2022 to examine the correlations between thirteen individual county-level characteristics and opioid-related mortality. Mortality data were sourced from the CDC WONDER database, which included many suppressed values. To overcome this limitation, we developed an effective imputation method to ensure a more complete and accurate analysis of the relationships between these area-based measures and opioid-related mortality.

To identify which county-level characteristics are most strongly correlated with opioid-related mortality, we leverage insights from exploratory data analysis alongside feature importance results derived from machine learning models. While machine learning excels at identifying predictive variables, its results can sometimes highlight relationships that are not meaningful in real-world contexts. Conversely, data analysis uncovers empirical relationships but may fall short in predictive ability. Integrating these two methods allows us to address the limitations of each and gain a more comprehensive understanding.

While traditional regression models are well-suited for estimating marginal effect sizes under assumptions of linearity and additivity, they are limited in capturing complex, nonlinear interactions that may drive opioid-related mortality. In contrast, machine learning approaches can model these complexities without prespecifying a functional form. This study prioritizes the latter capability, as our goal is to uncover potentially nonlinear patterns across diverse county contexts rather than restrict our analysis to relationships that can be expressed in a simple linear framework. Additionally, machine learning and artificial intelligence approaches have been successfully applied in medicine to detect patterns and uncover relationships \cite{medicine_ai_1, medicine_ai_2, medicine_ml_1, medicine_ml_2}, further motivating their use here.

We begin this study by examining how the rates of each characteristic manifest in counties with anomalously high and low mortality rates. By identifying variables that exhibit their highest levels in high-mortality counties and their lowest levels in low-mortality counties, we highlight factors with bidirectional impacts across the mortality spectrum. This exploratory data analysis provides interpretable, preliminary insights into variable relationships before applying machine learning methods. We then identify which characteristics are key contributors to opioid-related mortality predictions using feature importance metrics from two machine learning models.

The first model, XGBoost, is a decision tree-based algorithm with a proven track record in classification and regression tasks across various domains \cite{xgb_og_paper,xgb_app_1,xgb_app_2,xgb_app_3}. It evaluates feature importance using information gain, which measures how much each feature improves predictive accuracy when splitting a decision tree node. The second model, an augmented autoencoder, compresses input data into a lower-dimensional latent space then decodes it back into the county-level space to produce mortality rate predictions \cite{auto_survey,auto_app_1,auto_app_2}. To tailor the architecture for this study, the autoencoder was augmented with a preliminary convolutional layer to aggregate the thirteen SVI variables. To quantify each feature's importance in the autoencoder, we utilized a Shapley Gradient Explainer, which identifies the variables that most significantly influence predictions when perturbed \cite{shap_gradient_explainer,integrated_gradients_paper,SHAP_values_paper}.

By integrating these complementary approaches, this study advances the understanding of the complex relationships between county-level characteristics and opioid-related mortality. Our findings reveal critical structural factors that are strongly correlated with opioid-related mortality, highlighting variables that may exacerbate the epidemic at high levels and mitigate it at low levels. These insights offer a valuable foundation for designing public health interventions aimed at addressing the ongoing opioid crisis.

\section{Materials and methods}\label{materials_and_methods}

\subsection{Mortality data}

This study utilizes county-level opioid-related mortality rates, measured per 100,000 persons, from 2010 to 2022. Throughout the study period, the county structure of the United States changed due to the formation of new counties and the reconfiguration of existing ones \cite{county_changes}. For example, Connecticut completely restructured its county boundaries in 2022 \cite{ct_2022_changes}. Such changes impacted data consistency across years. Also, the machine learning models used in this study require a fixed input dimension; therefore the data has been curated to reflect the 2022 county structure, which includes 3,144 counties. This step ensures that we provide the most up-to-date and cohesive representation of the national county landscape with respect to the data. 

This mortality data was sourced from the CDC WONDER online database. The data were pulled using underlying cause-of-death codes from the Tenth Revision of ICD (ICD–10): X40–X44 (unintentional), X60–X64 (suicide), X85 (homicide), and Y10–Y14 (undetermined) \cite{icd_codes_general,icd_codes_specific}. It is therefore important to note that these rates serve as indicators of opioid misuse rather than exact counts of opioid overdose mortality. These data were retrieved at the county level, with the specified ICD-10 codes, while the other categories were left in their default settings.

\subsection{Method for handling missing data}\label{missing_data_handling_section}

The CDC WONDER database suppresses death counts in counties with fewer than 10 deaths. As a result, many mortality rates were missing, about half the nation's counties in each year of the study period. Thus to use this data in our study, it was essential to develop an effective methodology for imputing the missing values. One approach used in a previous study involved imputing missing death rates utilizing available data from neighboring counties \cite{missing_wonder_rates}. In their method, each county's neighborhood was defined as the entire state in which that county resides. In this study, we adopt a similar approach but use a more localized strategy: We impute missing mortality rates by averaging the available rates from a county’s queen adjacent neighbors—counties that share a direct border or a corner. 

Utilizing the queen adjacent counties, our methodology proceeds in a step-wise manner. First, rates for counties with exactly one missing neighbor are imputed by averaging the available neighboring county rates. Next, counties with two missing neighbors are addressed in the same way, followed by those with three missing neighbors. This process repeats iteratively until only counties for which all neighbors have available data remain. These counties are handled at the end because they do not impact the earlier steps, and their missing rates are imputed by averaging over the full set of neighboring counties.

Next, we address island counties with missing data. These counties are geographically separated from the mainland and therefore lack queen adjacent neighbors. For these island counties, which number at most three in any given year, we construct neighborhoods using the closest continental neighbors. Missing rates are then imputed by averaging rates over these constructed continental neighborhoods.

Our approach focuses exclusively on imputing mortality rates rather than raw death counts. This distinction is critical because mortality rates are calculated relative to a county’s population; small changes in raw death counts can lead to disproportionately large shifts in mortality rates, particularly in counties with small populations. Imputing raw death counts could introduce significant variability and potentially distort the derived rates, undermining the reliability of the imputation. By focusing directly on the mortality rates, our approach minimizes sensitivity to population size and ensures more consistency in the data.

To evaluate the effectiveness of our imputation strategy, we conducted an efficacy comparison using a complete national-level dataset, spanning 2014 to 2020, of opioid-related mortality rates sourced from HepVu \cite{hepvu_rates}. To simulate the level of missingness present in our study data, we randomly censored approximately half the county rates in each year. This allowed us to assess the efficacy of various imputation techniques by comparing each method’s imputed values to the original, uncensored rates.

We compared our method against three alternative strategies. The first was national mean imputation, where missing values were imputed using the national average of all available county rates. The second was state mean imputation \cite{missing_wonder_rates}, where missing values were imputed using the average of available county rates within the same state. This approach suffered, however, in years where entire states had no uncensored counties, in which case we defaulted to the national mean. The third was inverse distance weighting (IDW), where for each missing county, available neighbors were weighted by the inverse of their geodesic distance from the target county’s centroid. Widely used in hydrology for estimating rainfall \cite{idw_rainfall_1, idw_rainfall_2}, IDW assigns greater influence to nearby locations and is a long-established method for geospatial interpolation.

The results of this comparison are presented in \cref{table_imputation_comparison}, with additional histograms available in the project’s GitHub repository. Performance was evaluated using mean absolute error (MAE), root mean squared error (RMSE), mean percentage error (MPE), and mean absolute percentage error (MAPE). Across all years, our method substantially outperformed both the national and state-level imputations. As seen in \cref{table_imputation_comparison}, while the IDW method achieved marginally lower errors in some years, our novel approach performed nearly identically. Beyond its strong performance, our neighbor-based method integrates more naturally into machine learning and deep learning workflows, yielding reproducible feature values without the added complexity of distance calculations or parameter tuning. Additionally, its simplicity, interpretability, and near-parity with IDW make it a highly practical choice; particularly in geospatial contexts where adjacency is both intuitive and meaningful.

\begin{table}[h!]
\centering
\begin{tabular}{
    |p{1cm}| 
    p{5cm}|
    p{1cm}|
    p{1cm}|
    p{1.5cm}|
    p{1cm}|
}
 \hline
\multicolumn{1}{|>{\columncolor{lightgrey}}c|}{\textbf{Year}} & 
\multicolumn{1}{>{\columncolor{lightgrey}}c|}{\textbf{Method}} & 
\multicolumn{1}{>{\columncolor{lightgrey}}c|}{\textbf{MAE}} & 
\multicolumn{1}{>{\columncolor{lightgrey}}c|}{\textbf{RMSE}} & 
\multicolumn{1}{>{\columncolor{lightgrey}}c|}{\textbf{MPE}} & 
\multicolumn{1}{>{\columncolor{lightgrey}}c|}{\textbf{MAPE}} \\
 \hline\noalign{\vskip 3.5pt}\hline

 \hline          %method      MAE    RMSE    MPE       MAPE
 2014            & National mean & 5.67 & 7.71 & -34.09\% & 55.48\%  \\
                 & State mean & 4.37 & 6.30 & -16.19\% & 35.03\%  \\
                 & IDW weighted mean & 3.35 & 4.83 & -9.21\% & 25.87\% \\
                 & Adjacent neighbors mean & 3.38 & 4.91 & -9.70\% & 26.14\% \\
\hline
 
\hline\noalign{\vskip 3.5pt}

 \hline          %method      MAE    RMSE    MPE       MAPE
 2015            & National mean & 6.23 & 8.40 & -30.53\% & 54.12\%  \\
                 & State mean & 4.63 & 6.56 & -13.70\% & 33.92\%  \\
                 & IDW weighted mean & 3.69 & 5.28 & -7.66\% & 25.54\% \\
                 & Adjacent neighbors mean & 3.71 & 5.28 & -8.03\% & 25.72\% \\
 \hline

\hline\noalign{\vskip 3.5pt}

 \hline          %method      MAE    RMSE    MPE       MAPE
 2016            & National mean & 7.54 & 10.24 & -35.14\% & 57.98\%  \\
                 & State mean & 5.34 & 7.73 & -15.58\% & 34.45\%  \\
                 & IDW weighted mean & 4.21 & 6.41 & -7.71\% & 24.93\% \\
                 & Adjacent neighbors mean & 4.20 & 6.38 & -7.91\% & 24.98\% \\
 \hline

 \hline\noalign{\vskip 3.5pt}

 \hline          %method      MAE    RMSE    MPE       MAPE
 2017            & National mean & 8.20 & 11.17 & -35.87\% & 58.90\%  \\
                 & State mean & 5.73 & 8.41 & -14.71\% & 34.20\%  \\
                 & IDW weighted mean & 4.54 & 6.78 & -8.11\% & 25.63\% \\
                 & Adjacent neighbors mean & 4.53 & 6.78 & -8.19\% & 25.67\% \\
 \hline

  \hline\noalign{\vskip 3.5pt}

 \hline          %method      MAE    RMSE    MPE       MAPE
 2018            & National mean & 7.83 & 10.43 & -36.94\% & 59.65\%  \\
                 & State mean & 5.46 & 7.82 & -15.26\% & 34.20\%  \\
                 & IDW weighted mean & 4.33 & 6.26 & -9.03\% & 25.90\% \\
                 & Adjacent neighbors mean & 4.33 & 6.26 & -9.49\% & 26.08\% \\
 \hline

   \hline\noalign{\vskip 3.5pt}

 \hline          %method      MAE    RMSE    MPE       MAPE
 2019            & National mean & 7.82 & 10.97 & -30.29\% & 54.67\%  \\
                 & State mean & 5.84 & 8.61 & -14.37\% & 34.62\%  \\
                 & IDW weighted mean & 4.59 & 7.03 & -5.73\% & 25.23\% \\
                 & Adjacent neighbors mean & 4.60 & 7.00 & -6.00\% & 25.35\% \\
 \hline

    \hline\noalign{\vskip 3.5pt}

 \hline          %method      MAE    RMSE    MPE       MAPE
 2020            & National mean & 10.37 & 15.01 & -32.20\% & 55.67\%  \\
                 & State mean & 7.60 & 11.65 & -13.08\% & 34.01\%  \\
                 & IDW weighted mean & 6.08 & 9.55 & -5.45\% & 25.49\% \\
                 & Adjacent neighbors mean & 6.09 & 9.56 & -5.70\% & 25.58\% \\
 \hline
\end{tabular}
\caption{Performance comparison of four mortality rate imputation methods. The global and state mean methods use national and state-level averages, respectively. The IDW weighted method imputes rates using inverse distance weighting of nearby counties, while the adjacent neighbors method (our novel approach) imputes using the average of available queen-adjacent neighbors. Metrics include mean absolute error (MAE), root mean squared error (RMSE), mean percentage error (MPE), and mean absolute percentage error (MAPE).}
\label{table_imputation_comparison}
\end{table}

\begin{figure*}[h!]
    \centering
    \includegraphics[width=\linewidth]{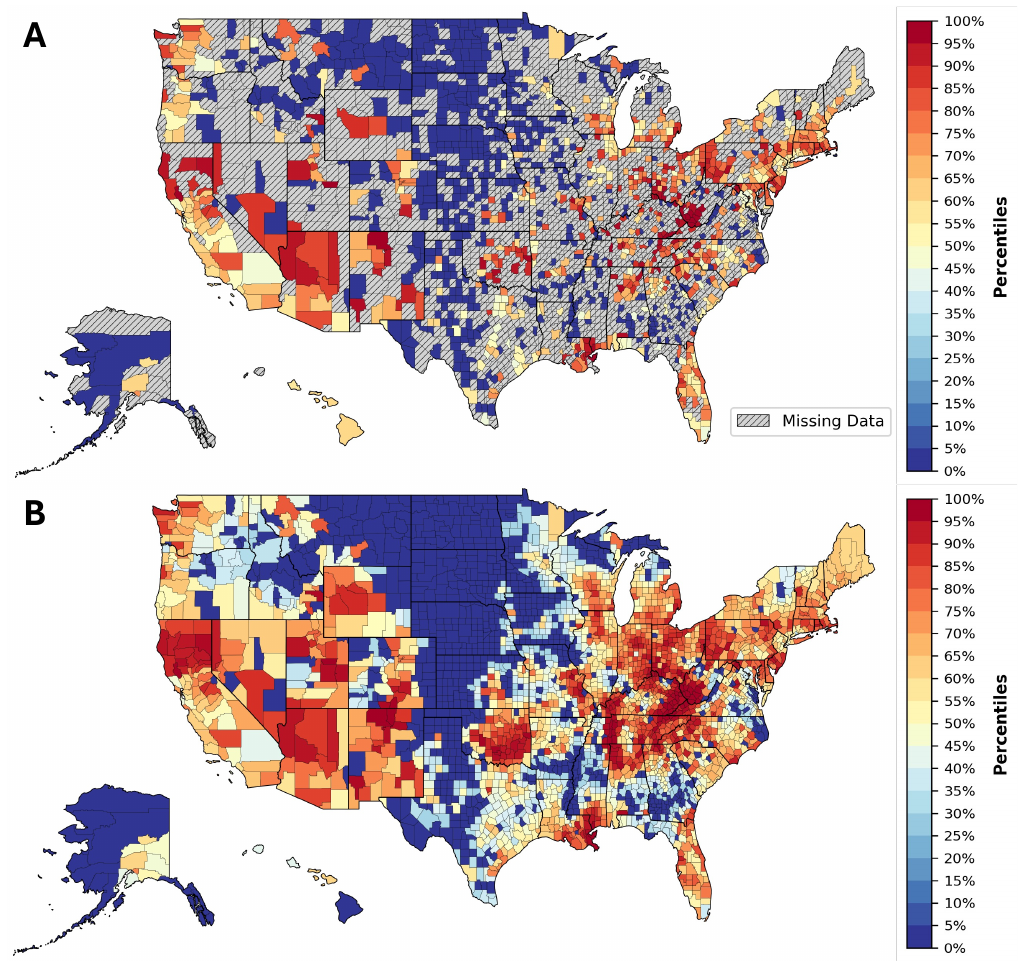} 
    \caption{Missing and interpolated data maps for the 2013 mortality rates. (\textbf{A}) 2013 missing data map. (\textbf{B}) 2013 interpolated data map. Each map is colored by percentiles from the respective empirical distributions. Figures are best viewed in color.}
    \label{fig_missing_data_maps}
\end{figure*}

To showcase the results of our methodology, we present national mortality rate maps for 2013. This year had the most significant data censoring, with 1,681 out of 3,144 counties lacking mortality rates, or 53.47\% of the data. \Cref{fig_missing_data_maps} displays maps both with and without the missing data, each colored by percentiles from the respective empirical distributions. \Cref{fig_missing_data_maps}a illustrates the extent of the missing data, revealing an incomplete view of national mortality rates. In contrast, \cref{fig_missing_data_maps}b shows the results after applying our imputation method; previously missing regions are now filled with interpolated data that preserves the spatial patterns of the original dataset. This significant improvement highlights the effectiveness of our approach, allowing us to address substantial data gaps and create a more complete and accurate national dataset for analysis.

Finally, Connecticut's county structure underwent a complete change in 2022. Since our dataset is curated to reflect the 2022 county boundaries, we needed to account for this structural change in all prior years of the study. To do this, we used the Python package Tobler \cite{tobler_github}, which specializes in spatial interpolation. Tobler allowed us to transfer data from the old county boundaries (2010 to 2021) to the new 2022 structure by interpolating mortality rates between the two geographic frameworks. This interpolation adjusts for differences in county areas, and ensures that population-dependent variables like mortality rates are accurately represented in the new structure.

\begin{figure*}[h]
    \centering
    \includegraphics[width=\linewidth]{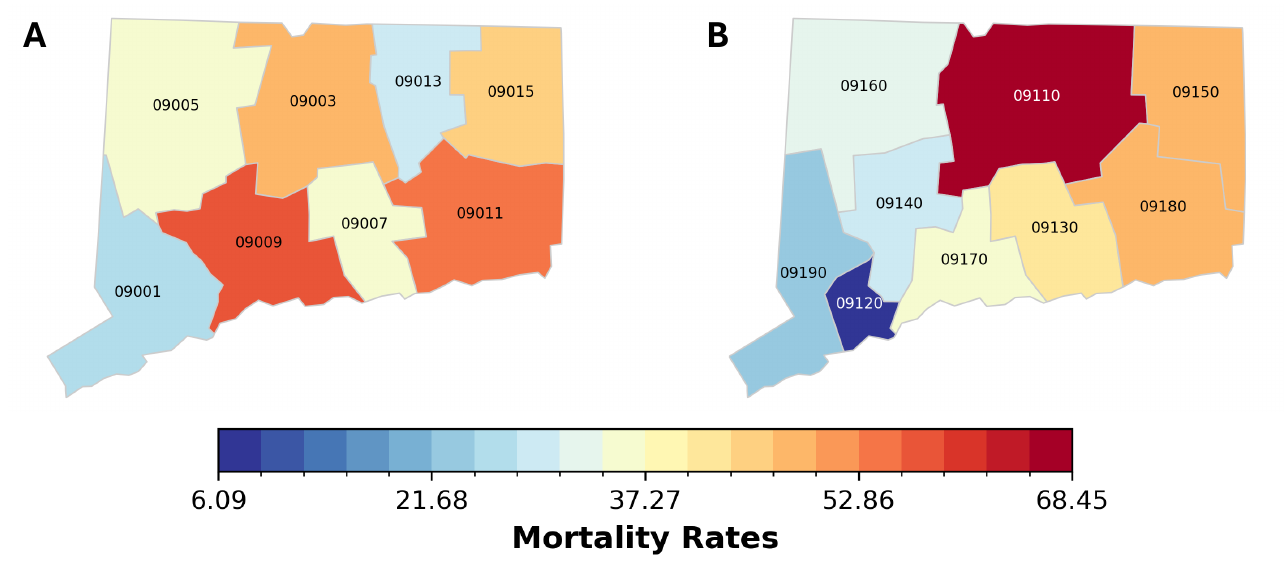} 
    \caption{Comparison of the 2021 mortality rates using the old and new Connecticut county structures. (\textbf{A}) 2021 mortality rates for the old (pre-2022) Connecticut county structure. (\textbf{B}) 2021 mortality rates for the new (2022) Connecticut county structure. The color scales for both maps are normalized to the minimum and maximum 2021 mortality rates across both county structures. The minimum 2021 mortality rate is 6.09, shown in the darkest blue, while the maximum rate is 68.45, shown in the darkest red. Figures are best viewed in color.}
    \label{fig_ct_interpolation}
\end{figure*}

\Cref{fig_ct_interpolation} demonstrates the results of using the Tobler library to interpolate mortality rates for Connecticut in 2021, the last year before the county structure changed. \Cref{fig_ct_interpolation}a displays Connecticut’s old county structure (pre-2022), while \cref{fig_ct_interpolation}b shows the new 2022 boundaries. For direct comparison, the color scales for both maps are normalized to the minimum and maximum 2021 mortality rates across both county structures. As an example, counties 09003 and 09013 from the old structure roughly correspond to county 09110 in the new 2022 structure. Here the interpolated data accurately reflects this consolidation, with the interpolated value being substantially larger than both old values. For another example, county 09009 in the old structure roughly corresponds to counties 09140 and 09170 in the new structure. In this example, the interpolated data captures this division effectively, with both interpolated values considerably lower than the old value.

\subsection{County-level characteristic data}

This study uses county-level rank ordered percentile estimates (ranging from 0 to 100) of thirteen SVI variables across the United States from 2010 to 2022. The set of variables utilized for 2010 data collection differs from those used in 2022. Therefore, we were forced to exclude all factors that were not available across every year in the study period, resulting in the thirteen characteristics utilized in this study. All data were sourced from the CDC and the Agency for Toxic Substances and Disease Registry's Social Vulnerability Index, which utilizes data from the American Community Survey to assess the resilience of communities to external stresses on human health \cite{svi_rates}. 

The 2010 data was not followed by another release until 2014, after which the data has been released biennially. To account for the gap between 2010 and 2014, we imputed the intervening years by calculating a quarter of the difference between the 2010 and 2014 data points and adding it consecutively to each year starting in 2014. After 2014, for the biennial data releases, we applied a similar method by calculating half the difference between consecutive data points and adding it to the earlier year’s rates. This method, with the understanding that yearly rate fluctuations are modest, provided a viable way to analyze the impact of these county-level characteristics on the national opioid crisis throughout the entire study period. 

Notably, Rio Arriba County, New Mexico, experienced a data collection error in 2018, resulting in a suppressed value for that year \cite{rio_arriba_data_error}. Therefore, in this situation, we applied the same approach used for the gap between 2010 and 2014, taking a quarter of the difference between the 2016 and 2020 data points and adding it consecutively to each year starting in 2016. Overall, the SVI data had minimal missing values, at most four counties in any given year of the study. These missing rates were imputed using the exact same methodology employed for the mortality rates. 

\subsection{XGBoost model}\label{xgboost_model}

In this study, XGBoost was used to predict opioid-related mortality rates from 2011 to 2022 based on the previous year’s data. To use XGBoost, several parameters need to be selected, such as the number and depth of trees. To select these parameters, we employed a grid search algorithm with cross-validation, optimizing for the highest predictive accuracy. For training and testing, we used 5-fold cross-validation to randomly split the input data into five folds. In each fold, XGBoost trained on four subsets and predicted on one. This process was repeated across all five folds to generate predictions for every county, ensuring that the model was both trained and tested on the entire dataset.

To describe this model mathematically, fix a year $t\in\{2010,2011,\dots,2021\}$ and let $X^{(t)}=\big[x_{i,j}^{(t)}\big]\in\mathbb{R}^{3144\times13}$ be the matrix of county-level characteristics input to the model in year $t$. Each row of matrix $X^{(t)}$ corresponds to a county, while each column corresponds to a characteristic. In particular, let $x_{i,*}^{(t)}$ denote the row vector containing all the feature data for county $i$, and let $x_{*,j}^{(t)}$ denote the column vector containing all county rates for variable $j$. Letting $Y^{(t+1)}=\big(y_i^{(t+1)}\big)\in\mathbb{R}^{3144}$ be the target vector of mortality rates in year $t+1$, we can define the full dataset $X^{(t)}\times Y^{(t+1)}$ passing through the model to be the set of all counties' feature vectors together with their corresponding target values: 
\begin{align*}
   X^{(t)}\times Y^{(t+1)}=\big\{\big(x_{i,*}^{(t)},y_i^{(t+1)}\big)\,\big|\,i\in\{1,2,\dots,3144\}\big\} 
\end{align*}

Starting with the root node, an individual decision tree grows by recursively applying a splitting process to each node $n$ in the tree. Let the data at node $n$ be denoted by $Q_n$ with $s_n$ samples. Then $s_n$ is the number of counties which reached that node, and $Q_n\subseteq X^{(t)}\times Y^{(t+1)}$ is the subset of data for those counties. At each node, a potential split $\theta=(j,h_n)$, consisting of a feature $j$ and a threshold $h_n$, partition $Q_n$ into two subsets:
\begin{align*}
    Q_n^{left}(\theta)&=\{(x_{i,*}^{(t)},y_i^{(t+1)})\,|\,x_{i,j}^{(t)}\le h_n\},\\
    Q_n^{right}(\theta)&=Q_n\setminus Q_n^{left}(\theta).
\end{align*}

$Q_n^{left}(\theta)$ is the data for counties satisfying the boolean condition of the split, meaning these data points proceed down the tree's left path; while $Q_n^{right}(\theta)$ contains the data for counties which did not satisfy the split condition, meaning these data points proceed down the right path. The quality of a potential split at node $n$ is measured using a loss function $L(Q)$ which evaluates how well the data in an arbitrary set $Q$ fits the target values. In particular, we utilize the mean absolute error as the loss function:
\begin{align*}
    L(Q)=\frac{1}{|Q|}\sum_{(x_{i,*}^{(t)}\,,y_i^{(t+1)})\in Q}|y_i^{(t+1)}-\hat{y}_i^{(t+1)}|,
\end{align*}
where $|Q|$ represents the total number of elements in the set $Q$, and $\hat{y}_i^{(t+1)}$ is the predicted mortality rate for county $i$. 

The gain $G(Q_n,\theta)$ from splitting $Q_n$ by $\theta$ is then given by
\begin{equation}\label{eq_gain}
    G(Q_n,\theta)=L(Q_n)-\bigg(\frac{s_n^{left}}{s_n}L\big(Q_n^{left}(\theta)\big)+\frac{s_n^{right}}{s_n}L\big(Q_n^{right}(\theta)\big)\bigg).
\end{equation}
Here $s_n^{left}$ is the number of counties which satisfied the split condition, and $s_n^{right}$ is the number that did not. The optimal split $\tilde{\theta}$ at this node is then chosen by maximizing the gain:
\begin{equation}\label{eq_argmax}
    \tilde{\theta}=\text{argmax}_\theta G(Q_n,\theta).
\end{equation}

This process creates child nodes from the subsets $Q_n^{left}(\tilde{\theta})$ and $Q_n^{right}(\tilde{\theta})$. It continues recursively until a stopping criterion is met, such as reaching the minimum number of samples at a node or when no further improvement in gain is possible. Mortality rate predictions are then made at these nodes with no further splits, called leaf nodes or terminal nodes. If $n$ is a terminal node, then the prediction for each county $i$ which reached this node is
\begin{align*}
    \hat{y}_{i}^{(t+1)}=\frac{1}{s_n}\sum_{(x_{i,*}^{(t)}\,,y_i^{(t+1)})\in Q_n}y_i^{(t+1)}.
\end{align*}

To evaluate the importance of each county-level characteristic in the model, XGBoost utilizes the gain from \cref{eq_gain}. This metric reflects the improvement in the model’s predictive performance when a county-level characteristic is used to split a decision node. At each node, the optimal split was chosen via \cref{eq_argmax}, which can be rewritten as follows:
\begin{equation}\label{eq_new_argmax}
    (\tilde{j},\tilde{h}_n)=\text{argmax}_{j,h_n} G\big(Q_n, (j,h_n)\big).
\end{equation}
From \cref{eq_new_argmax}, we see that at each node, XGBoost evaluates all features $j\in\{1,2,\cdots,13\}$ and selects the optimal feature $\tilde{j}$, along with a corresponding threshold, to maximize the gain.

As the decision trees are built, XGBoost sums the gains for each feature across all nodes and trees in the model. The total gain for feature $j$, denoted $G_{total}(j)$, is then the sum of the gains $G\big(Q_n,(\tilde{j},\tilde{h}_n)\big)$ over all nodes $n$ in the model which split using feature $j$:
\begin{align*}
    G_{total}(j)=\sum_{n=1}^J G\big(Q_n,(\tilde{j},\tilde{h}_n)\big).
\end{align*}
Here $J$ is the number of nodes which used feature $j$ to split. The features that contribute the most to improving the model’s predictive accuracy are those with the highest total gain. Consequently, variables with higher total gain are considered more important, as they most consistently reduce the prediction error across the model. 

\subsection{Autoencoder model}

In this study, the autoencoder model was used to predict opioid-related mortality rates from 2011 to 2022 based on the previous year’s data. The input data is first passed through the initial 1D convolutional layer, which aggregates the thirteen characteristics before being fed into the encoder. The encoder compresses the data into a lower-dimensional representation, which is then passed to the decoder to produce the final vector containing yearly mortality rate predictions for each county in the nation. 

To preserve the natural temporal structure of the opioid crisis, the model is trained sequentially on data from 2010 to 2020, with 2015 held out as the validation set, and 2021 used for testing. To prevent overfitting, the model is allowed to train for up to 100 epochs, with early stopping triggered after 10 consecutive worse predictions on the validation set. The training process utilizes an $L^1$ loss function and a cyclical learning rate.

To describe the autoencoder model mathematically, again let $t\in\{2010, 2011, \dots, 2021\}$ denote a fixed year, and $X^{(t)}$ denote the input data from year $t$. Let $\hat{Y}^{(t+1)}$ be the output vector of mortality rate predictions for year $t+1$, and let $A$ be the full neural network model. It consists of five layers: the initial 1D convolutional layer $C$, two encoding layers $E_1$ and $E_2$, and two decoding layers $D_3$ and $D_4$. Then the model can be expressed as follows:
\begin{align*}
    \hat{Y}^{(t+1)} &= A\big(X^{(t)}\big)\\
    &= \big(D_4 \circ D_3 \circ E_2 \circ E_1 \circ C\big)\,\big(X^{(t)}\big).
\end{align*}

The convolutional layer $C$ learns an effective way to convolve the thirteen county-level characteristics, producing a single vector of size 3144. This vector represents the aggregate state of the nation with respect to the variables. The convolutional layer can be written as:
\begin{equation*}
    C\big(X^{(t)}\big) = W_C \cdot X^{(t)} + b_C,
\end{equation*}
where $W_C$ is the matrix of weights for this layer, and $b_C$ is the vector of biases, both optimized through a stochastic gradient descent process.

The encoder layers $E_1$ and $E_2$ iteratively reduce the dimensionality of the input data. $E_1$ is followed by a ReLU activation function, while $E_2$ is not. For an arbitrary input $x$, these layers can be written as follows:
\begin{align*}
    E_1(x) &= \text{ReLU}\big(W_{E_1} \cdot x + b_{E_1}\big),\\
    E_2(x) &= W_{E_2} \cdot x + b_{E_2}.
\end{align*}
Here $W_{E_i}$ and $b_{E_i}$ are the weight matrices and bias vectors for $i\in\{1,2\}$, respectively.

The decoder layers $D_3$ and $D_4$ then iteratively increase the dimensionality back to the original input space. For an arbitrary input $x$, these layers can be written as follows:
\begin{align*}
    D_3(x) &= \text{ReLU}\big(W_{D_3} \cdot x + b_{D_3}\big)\\
    D_4(x) &= W_{D_4} \cdot x + b_{D_4}.
\end{align*}
Here $W_{D_i}$ and $b_{D_i}$ are the weight matrices and bias vectors for $i\in\{3,4\}$, respectively.

To gain insight into the most important features in the autoencoder’s mortality predictions, we employed a Shapley gradient explainer. This method, building on the concept of integrated gradients \cite{integrated_gradients_paper}, utilizes the expected gradient approach to quantify the impact of each input feature \cite{expected_gradients}. Expected gradients are calculated by averaging the gradient of the model’s output with respect to each feature over the entire county structure, providing a measure of how each component impacts the autoencoder's predictions. Each variable’s impact is quantified via a SHAP value \cite{SHAP_values_paper}, representing the contribution of that feature to the predicted mortality rates. Consequently, the most important features in the autoencoder model are those for which minor perturbations produce the greatest changes in the model’s output.

\section{Results}

\subsection{Data analysis}

We sourced mortality rates from the Centers for Disease Control and Prevention's Wide-ranging ONline Data for Epidemiologic Research (CDC WONDER) system \cite{cdc_wonder}, which suppresses death counts in counties with fewer than 10 deaths. Consequently, mortality rates were unavailable for nearly half of the nation's counties in each year of the study period. To address this data censoring, we developed a novel methodology to impute missing values based on data from neighboring counties, as detailed in \cref{missing_data_handling_section}.

Using this imputed dataset, our study begins by analyzing how the rates of each SVI component manifests in counties with anomalous mortality rates. For each year in the study period, we define anomalies as rates that fall within specific percentiles of a statistical distribution fitted to the nonzero mortality rates. Directly including the zero values in the fitting process poses significant challenges. For example, zero values account for more than 30\% of the data in several years, creating a large spike at zero that distorts both parametric and empirical distributional fits.

To address these challenges, we analyzed counties with zero mortality rates separately from those with nonzero rates. We found that the characteristic patterns in the zero-rate counties and the anomalously low-rate counties were nearly identical. These findings will be discussed in greater detail later in this section, but we note here that these nearly indistinguishable patterns support our decision to focus on the distribution of nonzero rates. This approach allows us to define anomalous mortality while still capturing the broader social patterns present in the zero-rate counties.

\begin{figure*}[h]
    \centering
    \includegraphics[width=\linewidth]{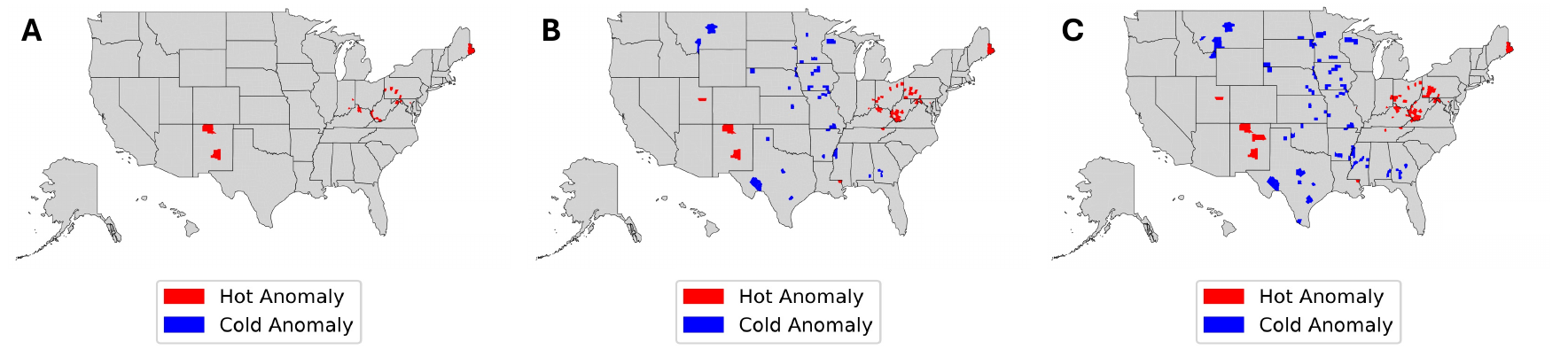} 
    \caption{Impact of threshold selection on 2016 anomalies. (\textbf{A}) 2016 anomaly map using 1\% tails. (\textbf{B}) 2016 anomaly map using 2\% tails. (\textbf{C}) 2016 anomaly map using 3\% tails. A 1\% threshold produced no cold anomalies, while a 2\% threshold provided the smallest bound that yielded cold anomalies. Increasing the threshold to 3\% further raised the number of anomalies in both categories. Figures are best viewed in color.}
    \label{fig_anomaly_tails}
\end{figure*}

Consequently, to define anomalous mortality rates, we restrict the set of candidate distributions to those defined on the domain $(0, \infty)$. To determine the best-fitting distribution, we evaluated Gamma, Weibull, inverse Gaussian, and lognormal distributions. The goodness-of-fit was assessed using the Kolmogorov-Smirnov test, Akaike Information Criterion, and Bayesian Information Criterion. Based on these statistical measures, the lognormal distribution consistently outperformed the other candidates, scoring as the best-fitting distribution each and every year in the study period. Using the fitted lognormal distributions, anomalies are defined annually in two categories as follows: ``Hot anomalies'' are counties with nonzero rates above the 98th percentile of the yearly fitted distribution, while ``cold anomalies'' are those whose nonzero rates fall below the 2nd percentile. To select the size of these tails, we assessed 1\%, 2\%, and 3\% thresholds. In some years, a 1\% tail produced no cold anomalies, whereas the 2\% tail always provided them. While a 3\% tail also could have been used, to maximize the anomalous nature of the rates, we opted for the tightest bounds that still yielded cold anomalies for analysis each year. \Cref{fig_anomaly_tails} shows how the anomalies in 2016 change with 1\%, 2\%, and 3\% tails.

\begin{figure*}[h]
    \centering
    \includegraphics[width=\linewidth]{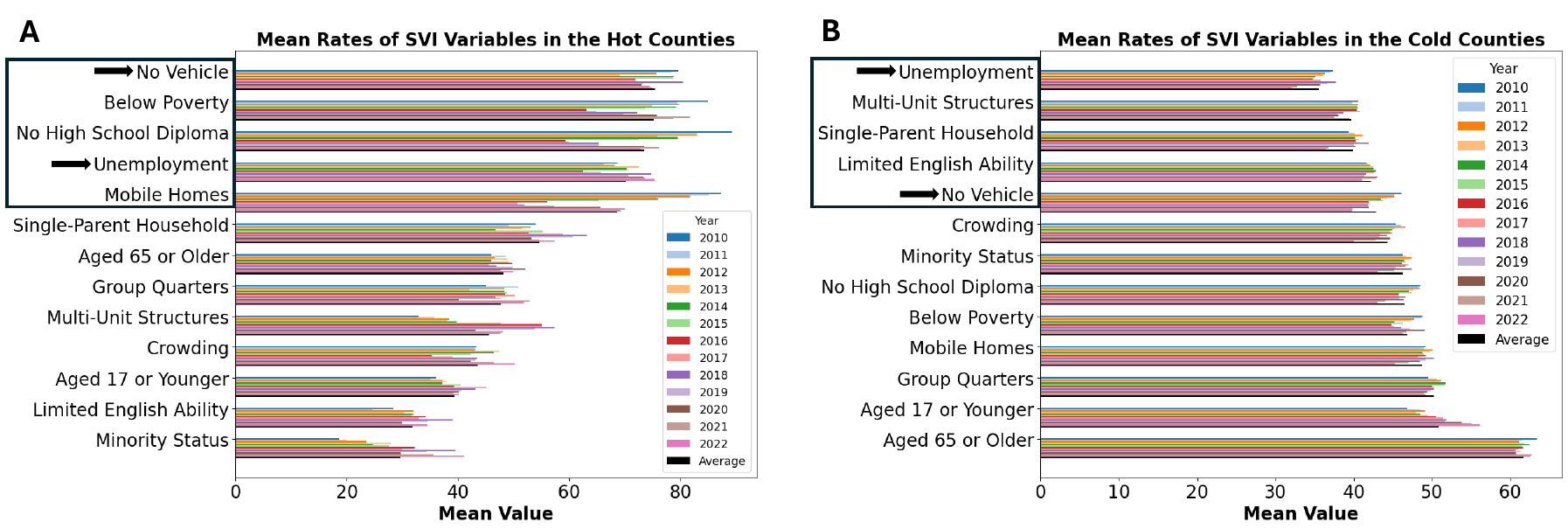}
    \caption{Mean rates of the county-level characteristics in the hot and cold anomalies. (\textbf{A}) Hot anomaly summary, where features are ranked from highest average mean rate down to lowest. (\textbf{B}) Cold anomaly summary, where features are ranked from lowest average mean rate down to highest. In both plots, each feature's yearly mean rates are displayed using colored bars, while the black bars showcase each feature's average mean rate across all years in the study. Figures are best viewed in color.}
    \label{fig_anomaly_summary}
\end{figure*}

To compare the rates of each characteristic within the anomalous counties, we rank the variables based on their mean rates. Specifically, we first calculate the mean rate within each year and then average these yearly means over the entire study period. Features are ordered from highest to lowest average mean rate in hot counties and from lowest to highest in cold counties. These rankings identify the variables most strongly correlated with anomalous mortality rates. The results are presented in figure \ref{fig_anomaly_summary}, where yearly mean rates for each feature are shown using colored bars, and black bars indicate each variable's average mean rate across all years of the study.

\Cref{fig_anomaly_summary}a ranks the characteristics based on their average mean rates in the hot counties. Among these hot counties, the variable with the highest average mean rate is no vehicle (75.40), followed by below poverty (75.22), no high school diploma (73.48), unemployment (70.18), and mobile homes (68.61). The high ranking of these features means they tend to manifest their highest rates in counties with anomalously high opioid-related mortality. This suggests that opioid-related mortality may be exacerbated by high levels of these factors. Beyond these top five, rates noticeably decline, starting with single-parent households (54.61) and continuing to drop further down the list. At the bottom are limited English ability and minority status, meaning they have the lowest average rates in the hot counties. This suggests that these characteristics may contribute the least to the exceptionally high mortality rates observed in the hot counties.

\Cref{fig_anomaly_summary}b ranks the variables based on their average mean rates in the cold counties. Among these cold counties, the characteristic with the lowest average mean rate is unemployment (35.58), followed by multi-unit structures (39.66), single-parent households (39.89), limited English ability (42.22), and no vehicle (42.84). The high ranking of these features means they tend to manifest their lowest rates in counties with anomalously low opioid-related mortality. This suggests that opioid-related mortality may be alleviated by lower levels of these factors. Beyond these top five, mean rates only gradually increase, except at the bottom, where a significant jump is observed for the variable aged 65 or older. This factor stands out with the highest average mean rate (61.74) in the cold anomalies, indicating that counties with anomalously low opioid-related mortality tend to have a significantly higher population of persons aged 65 or older.

\begin{figure*}[h]
    \centering
    \includegraphics[width=\linewidth]{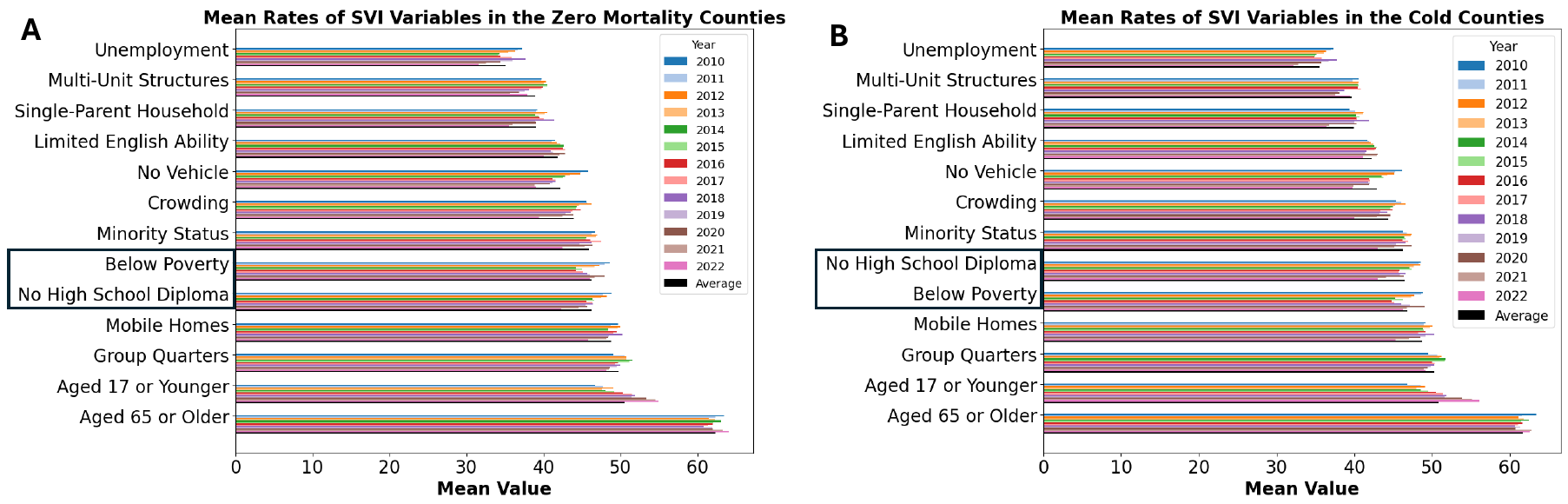}
    \caption{Mean rates of the county-level characteristics in the zero mortality and cold counties. (\textbf{A}) Zero mortality counties summary, where features are ranked from lowest average mean rate down to highest. (\textbf{B}) Cold counties summary, where features are also ranked from lowest average mean rate down to highest. In both plots, each feature's yearly mean rates are displayed using colored bars, while the black bars showcase each feature's average mean rate across all years in the study. Figures are best viewed in color.}
    \label{fig_frozen_cold}
\end{figure*}

As noted at the beginning of this section, the pattern observed in the zero mortality rate counties and the one just described for the cold counties is nearly indistinguishable. This is illustrated in \cref{fig_frozen_cold}, where the only difference is a swap in order between no high school diploma and below poverty, both of which have average means that were nearly the same.

In this data analysis, notably, only two variables---unemployment and lack of vehicle access---exhibited bidirectional relationships, with their highest rates occurring in counties with anomalously high mortality and their lowest rates in counties with anomalously low mortality. The fact that these features are observed at both extreme ends of the mortality rate spectrum suggests that they may play a particularly significant role: High levels may contribute to increased opioid-related mortality, while low levels may help reduce it. These empirical results provide a real-world foundation for the subsequent machine learning analysis, identifying which features are the key contributors to opioid-related mortality predictions.

\subsection{Efficacy of machine learning models}

This study employed two machine learning models to predict opioid-related mortality rates from 2011 to 2022 using the previous year’s data. XGBoost generates predictions for each county by combining the outputs from an ensemble of decision trees, where each tree iteratively corrects the errors of its predecessors through gradient-based optimization. The autoencoder model, on the other hand, processes input data through an initial 1D convolutional layer that aggregates the county-level characteristics before passing the data to an encoder, which compresses it into a lower-dimensional representation. This representation is then decoded to produce the yearly mortality rate predictions.

The efficacy of our machine learning models in forecasting county-level mortality rates is evaluated using both accuracy and error metrics. For each county, the error is measured as the absolute residual. To assess overall accuracy, the maximum absolute residual, which indicates the largest annual prediction error, is used as a benchmark: Each county’s error is normalized by this highest observed error across the country. The performance of the models is summarized in \cref{table_efficacy}, which presents the average errors, maximum errors, and average accuracy of both models for each predictive year in the study period; \cref{table_mort_data_stats} provides summary statistics for the mortality rate data to contextualize these results. 

\begin{table}[ht]
\centering
\begin{tabular}{c@{\hspace{30pt}}c@{\hspace{10pt}}c@{\hspace{30pt}}c@{\hspace{10pt}}c@{\hspace{30pt}}c@{\hspace{10pt}}c@{\hspace{30pt}}c}
\toprule
Year & \multicolumn{2}{l}{Avg Error} & \multicolumn{2}{l}{Max Error} & \multicolumn{2}{l}{Avg Accuracy} \\
     & XGB & AE & XGB & AE & XGB & AE \\
\midrule
2011 & 5.79 & 4.07 & 100.18 & 72.41 & 94.22\% & 94.39\% \\
2012 & 5.67 & 4.08 & 72.14 & 72.88 & 92.14\% & 94.40\% \\
2013 & 5.86 & 3.97 & 78.39 & 55.25 & 92.52\% & 92.82\% \\
2014 & 6.43 & 3.56 & 87.57 & 59.17 & 92.66\% & 93.98\% \\
2015 & 6.65 & 3.26 & 112.40 & 68.05 & 94.09\% & 95.22\% \\
2016 & 7.83 & 4.66 & 107.68 & 100.36 & 92.73\% & 95.36\% \\
2017 & 8.34 & 4.15 & 126.24 & 72.20 & 93.39\% & 94.25\% \\
2018 & 8.18 & 3.94 & 98.80 & 48.67 & 91.72\% & 91.91\% \\
2019 & 8.43 & 4.02 & 97.56 & 60.61 & 91.36\% & 93.36\% \\
2020 & 10.78 & 7.35 & 136.70 & 92.04 & 92.11\% & 92.02\% \\
2021 & 13.84 & 11.60 & 161.62 & 184.12 & 92.06\% & 93.70\% \\
2022 & 12.44 & 12.37 & 236.83 & 230.10 & 94.75\% & 94.62\% \\
\bottomrule
\end{tabular}
\caption{Efficacy metric summary for both the XGBoost (XGB) and autoencoder (AE) models. For each county, the error is measured as the absolute residual, while the accuracy is calculated as the county’s error normalized by the highest observed error nationwide.}
\label{table_efficacy}
\end{table}

\begin{table}[h!]
\centering
\begin{tabular}{c@{\hspace{30pt}}c@{\hspace{15pt}}c@{\hspace{15pt}}c@{\hspace{15pt}}c@{\hspace{15pt}}c@{\hspace{15pt}}c@{\hspace{15pt}}c}
\toprule
Year & Mean & Std Dev & Min & Q1 & Median & Q3 & Max \\
\midrule
2010 & 9.75 & 11.88 & 0.0 & 0.00 & 7.53 & 14.79 & 126.85 \\
2011 & 10.33 & 11.87 & 0.0 & 0.00 & 8.66 & 15.48 & 132.37 \\
2012 & 10.04 & 10.80 & 0.0 & 0.00 & 8.44 & 15.58 & 98.48 \\
2013 & 10.45 & 10.55 & 0.0 & 0.00 & 9.07 & 16.26 & 100.59 \\
2014 & 11.66 & 12.09 & 0.0 & 0.00 & 9.69 & 17.98 & 100.56 \\
2015 & 12.43 & 12.46 & 0.0 & 0.00 & 10.72 & 19.28 & 141.16 \\
2016 & 14.64 & 14.61 & 0.0 & 0.00 & 12.46 & 22.50 & 134.08 \\
2017 & 15.75 & 15.72 & 0.0 & 0.00 & 13.54 & 23.92 & 152.70 \\
2018 & 14.88 & 14.56 & 0.0 & 0.00 & 12.82 & 23.77 & 123.36 \\
2019 & 15.37 & 14.85 & 0.0 & 0.52 & 13.12 & 24.05 & 119.13 \\
2020 & 21.84 & 20.29 & 0.0 & 4.22 & 18.94 & 32.82 & 171.44 \\
2021 & 27.29 & 23.93 & 0.0 & 8.80 & 24.32 & 39.30 & 212.38 \\
2022 & 27.27 & 23.20 & 0.0 & 9.62 & 25.46 & 39.11 & 238.27 \\
\bottomrule
\end{tabular}
\caption{Summary statistics for the mortality rates from 2010 to 2022.}
\label{table_mort_data_stats}
\end{table}

\begin{figure*}[h]
    \centering
    \includegraphics[width=\linewidth]{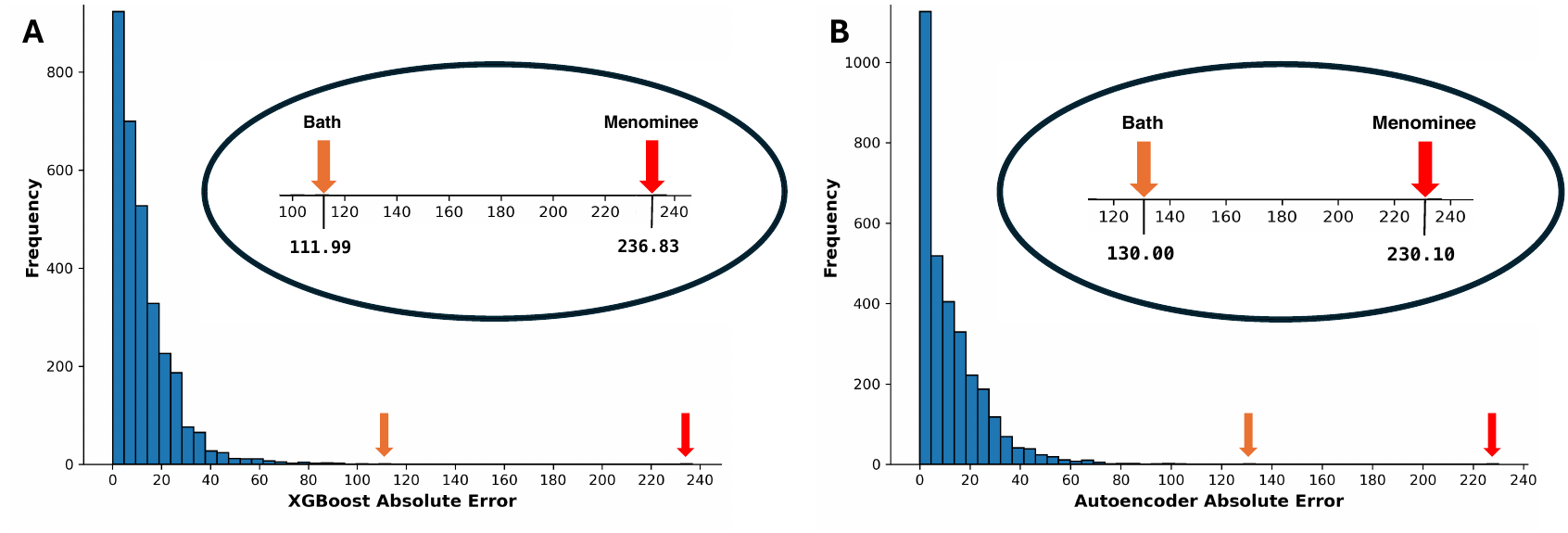} 
    \caption{Error histograms for 2022 mortality rate predictions from both models. (\textbf{A}) Histogram displaying the distribution of XGBoosts's absolute errors for the 2022 mortality rate predictions. (\textbf{B}) Histogram displaying the distribution of the autoencoder's absolute errors for the 2022 mortality rate predictions. In both plots, the error stemming from Menominee County, Wisconsin, is indicated by the red arrow, while the error stemming from Bath County, Kentucky, is indicated by the orange arrow. Figures are best viewed in color.}
    \label{fig_error_histos}
\end{figure*}

\begin{figure*}[h!]
    \centering
    \includegraphics[width=\linewidth]{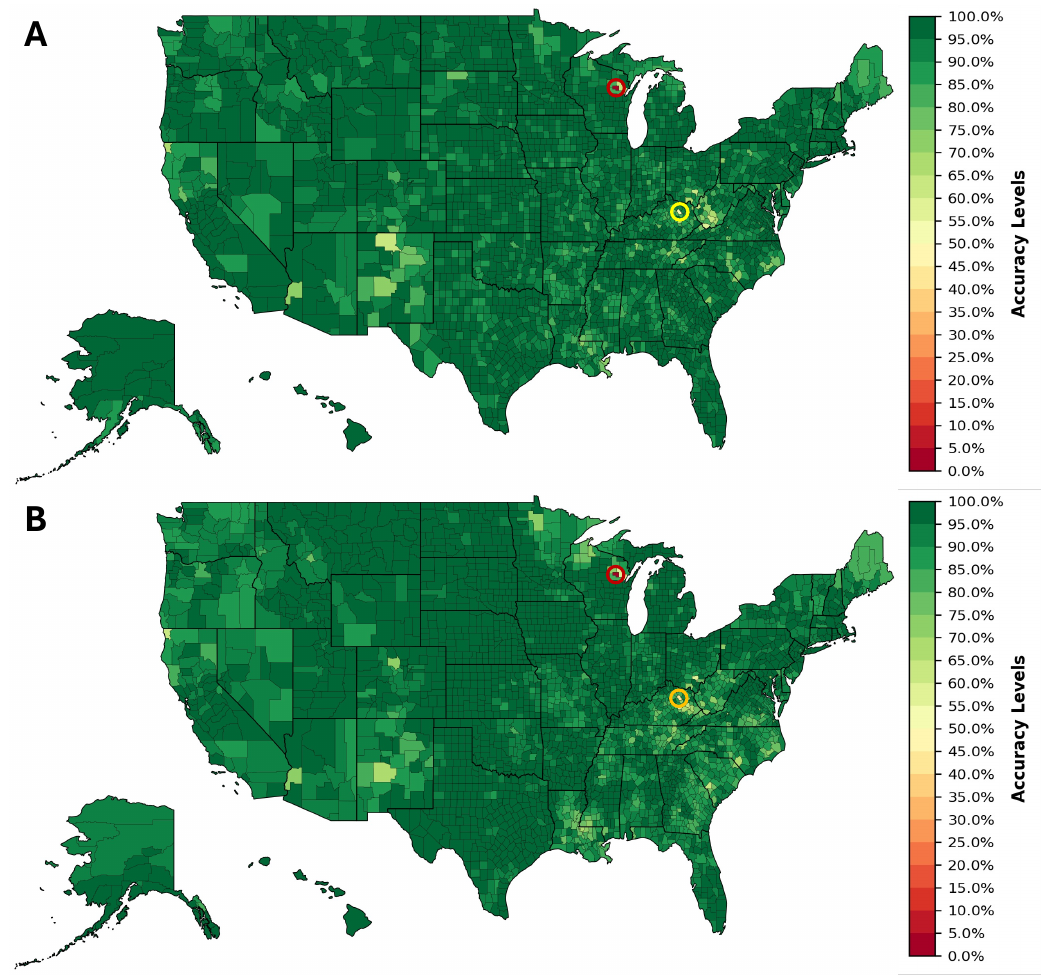} 
    \caption{Accuracy maps for 2022 mortality rate predictions from both models. (\textbf{A}) Accuracy map for XGBoost's 2022 mortality rate predictions. (\textbf{B}) Accuracy map for the autoencoder's 2022 mortality rate predictions. Both maps are color-coded to represent increasing intervals of accuracy by 5\%, starting from dark red for the least accurate predictions, and progressing to dark green for the most accurate predictions. In both plots, Menominee County, Wisconsin, which had the largest 2022 error for both models, is shaded in the darkest red. To make it easier to locate, this county is also circled in the same color. Bath County, Kentucky, is shaded in orange on the autoencoder’s accuracy map and in yellow on XGBoost’s map, highlighting that the autoencoder had a higher error here than XGBoost. This county has also been circled in its respective color for easier visual identification. Figures are best viewed in color.}
    \label{fig_accuracy_maps}
\end{figure*}

While each year is treated as an independent test year by XGBoost, 2022 serves as the sole testing year for the autoencoder. This approach was chosen to preserve the natural temporal trajectory of the opioid crisis by training the autoencoder sequentially on data from 2010 to 2020, with 2021 as the testing year and predictions made for 2022. Accordingly, we present visual representations of the 2022 efficacy results using error histograms (\cref{fig_error_histos}) and accuracy maps (\cref{fig_accuracy_maps}). 

As seen in \cref{table_efficacy}, both models generally maintain high accuracy throughout the study period, yet exhibit a notable increase in average and maximum errors starting in 2020 which persist through 2022. Some of these errors could be attributable to problems inherent in data censoring. The maximum error for both models in 2022 arises in Menominee County, Wisconsin, where the 2021 mortality rate was censored and subsequently imputed using our method. This imputation, based on neighboring counties, produced a value of 26.16; yet, the reported rate for 2022 was 238.27—a 810.82\% increase that neither model could predict. This extreme increase could be attributed to a reporting error in either 2021 or 2022, or to the low imputed value from the previous year due to the censored data in that year. Not all large errors might be attributable to data censoring however. For example, the second largest error in 2022 for both models occurs in Bath County, Kentucky. This county reported a mortality rate of 86.09 in 2021, which nearly doubled to 163.69 in 2022---a 90.14\% increase from the previous year that both models struggled to predict. The errors stemming from Menominee and Bath counties are highlighted in \cref{fig_error_histos}, which displays the complete 2022 error distributions for both models, providing a visual comparison of their error ranges.


\Cref{fig_accuracy_maps} shows each model's 2022 accuracy maps, both characterized predominantly by green hues, reflecting high predictive accuracy nationwide. Menominee County, Wisconsin, which had the largest 2022 error for both models, is marked correspondingly in the darkest shade of red. While Bath County, Kentucky, appears in orange on the autoencoder’s accuracy map and in yellow on XGBoost’s map. This discrepancy is due to the sharp increase in mortality rates in Bath County from 2021 to 2022, which contributed to a larger error for the autoencoder than XGBoost in that year.

\subsection{XGBoost feature importance}

We used XGBoost to predict opioid-related mortality rates from 2011 to 2022 based on the previous year’s data. To assess the importance of each county-level characteristic, XGboost utilizes information gain:
\begin{equation*}
    G(Q_n,\theta)=L(Q_n)-\bigg(\frac{s_n^{left}}{s_n}L\big(Q_n^{left}(\theta)\big)+\frac{s_n^{right}}{s_n}L\big(Q_n^{right}(\theta)\big)\bigg).
\end{equation*}
This metric, detailed in \cref{xgboost_model}, quantifies the improvement in the model’s predictive performance when a feature is used to split a node in a decision tree. As a result, the features with the highest gain scores are those that most significantly improve the model’s ability to predict opioid-related mortality rates.

\begin{figure*}[h]
    \centering
    \includegraphics[width=\linewidth]{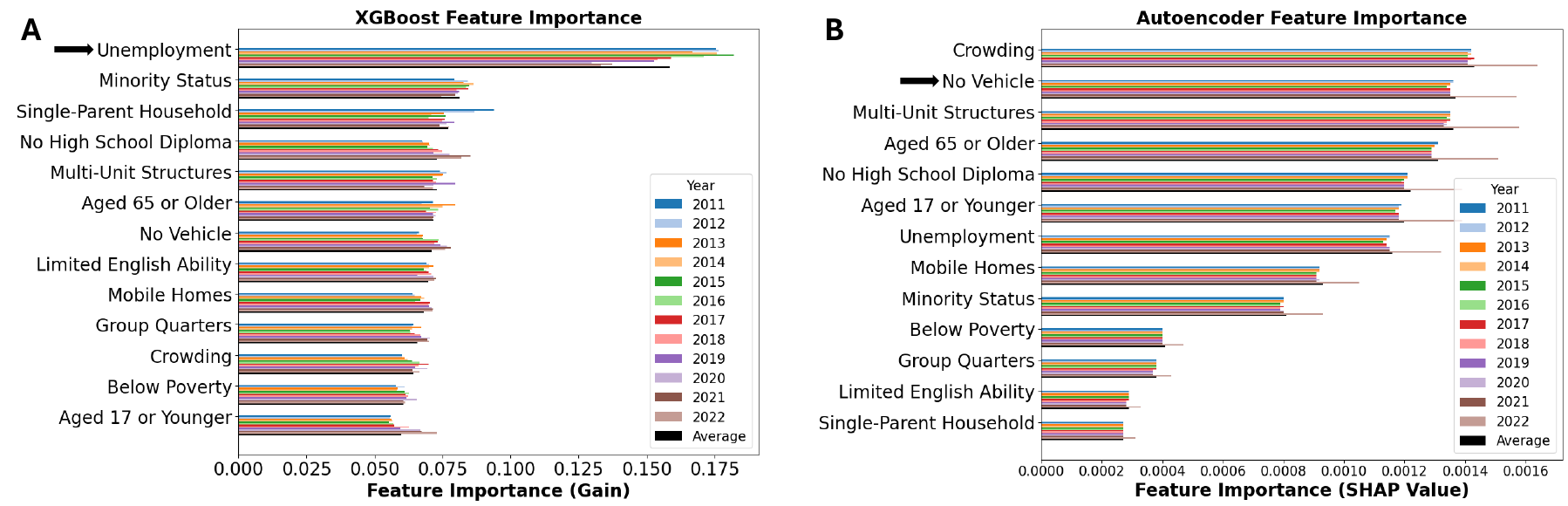} 
    \caption{Feature importance rankings for both models. (\textbf{A}) XGBoost feature importance summary. (\textbf{B}) Autoencoder feature importance summary. Both plots display each feature's yearly importance scores with colored bars, while black bars represent the average importance scores across all years. Features are ranked by their average importance, from highest to lowest in both plots. Figures are best viewed in color.}
    \label{fig_feature_importances}
\end{figure*}

The feature importance results for XGBoost are displayed in \cref{fig_feature_importances}a. In this image, each feature’s yearly importance scores are shown with colored bars, while each feature's average importance score across the entire study period is represented by the black bars. These black bars rank the components from highest average importance down to lowest. 

Significantly surpassing all others, unemployment (0.1585) has the highest average importance score. This is followed by minority status (0.0814), but its average importance score is only slightly over half that of unemployment. While its yearly importance scores fluctuate slightly, unemployment occupies the top spot every year. As we move further down the list, feature importance scores only continue to decrease. Notably, the dominant role unemployment plays in predicting opioid-related mortality rates for XGBoost corroborates the empirical findings of the data analysis.

\subsection{Autoencoder feature importance}

In this study, the autoencoder model was used to predict opioid-related mortality rates from 2011 to 2022 based on the previous year's data. To determine which county-level characteristics were most important in the autoencoder's predictions, we applied a Shapley Gradient Explainer to the initial convolutional layer. The Shapley Gradient Explainer generates SHAP (Shapley Additive Explanations) values, which quantify the contribution of each feature to a model's predictions \cite{shap_gradient_explainer,integrated_gradients_paper,SHAP_values_paper}. By measuring how sensitive the autoencoder's mortality predictions are to small changes in each feature, the SHAP values identify the most important components as those for which minor changes lead to the greatest impact on predictions.

The feature importance results for the autoencoder are showcased in \Cref{fig_feature_importances}b. In this image, each feature’s yearly SHAP value is shown with colored bars, while each feature's average SHAP value across the entire study period is represented by the black bars. These black bars rank the components from highest average importance down to lowest. 

The variable with the highest average SHAP value is crowding (0.00143), followed closely by no vehicle (0.00137) and multi-unit structures (0.00136). In fourth place is aged 65 or older (0.00131), from which SHAP values gradually decrease down to unemployment (0.00116) in seventh place. Beyond unemployment, there is a noticeable drop in the SHAP values for mobile homes and minority status. After which, there is another steep decline in importance moving down to the the last four characteristics. This reflects the much smaller contribution these four factors make to the model's predictions. Notably, the high importance score of no vehicle corroborates the empirical findings from the data analysis.

Lastly, we note that SHAP values are relative to the model, data, and problem context; they reflect the contributions of each feature to the predictions compared to the baseline (i.e., the average prediction). Ranking these values within the model provides a reliable and interpretable measure of feature importance, offering meaningful insights into the influence of each variable on this specific model and dataset.

\section{Discussion}

Utilizing thirteen county-level characteristics, this study sought to identify which variables are most strongly correlated with opioid-related mortality. The data analysis portion of this study examined how the rates of each component manifested in counties with anomalous mortality. Notably only two factors---unemployment and lack of vehicle access---exhibited bidirectional relationships, with their highest rates observed in counties with anomalously high mortality and their lowest rates in counties with anomalously low mortality. This pattern suggests that reducing unemployment and improving vehicle access may lead to measurable reductions in opioid-related mortality, particularly in the most severely affected regions.

These empirical results provided real-world grounding for the subsequent machine learning analysis, which identified the characteristics that are key contributors to opioid-related mortality predictions. This portion of the study employed two machine learning models: XGBoost and an autoencoder. Feature importance was assessed using information gain for XGBoost and a Shapley gradient explainer for the autoencoder. While the feature importance rankings differed between the two models, this difference highlights the value of employing multiple approaches, as it revealed critical insights that might have been overlooked if only one model had been used.

In the XGBoost model, unemployment emerged as the most important feature by a significant margin; the second-ranked feature, minority status, had an importance score just over half that of unemployment. This indicates that unemployment is nearly twice as strong a predictor of opioid-related mortality than any other variable in the XGBoost model. Although crowding ranked highest in the Shapley analysis of the autoencoder, lack of vehicle access followed closely in second place with a nearly identical importance score. This indicates that small changes in vehicle access rates produced substantial shifts in the autoencoder’s mortality predictions. These results from the machine learning analysis corroborate those from the data analysis; they underscore unemployment and lack of vehicle access as strongly correlated with opioid-related mortality and highlight their potential roles as key factors in the opioid epidemic. 

By quantifying the contribution of unemployment to opioid-related mortality predictions using feature importance metrics from two machine learning models, this analysis builds on the well-established connection between unemployment and opioid-related outcomes \cite{macroeconomic_report,ucla_nursing_unemployment,unemployment_and_oud}. Previous work has shown that unemployment, along with the accompanying economic hardships, significantly contributes to opioid misuse in the United States \cite{unemployment_association,economic_hardship,aspe_paper}. These studies emphasize the importance of employment-related factors in addressing the opioid crisis, and call for substance abuse treatment to remain a priority even during economic downturns. Beyond economic strain, unemployment has been shown to disrupt the structure, routine, social connections, and sense of identity that are critical to recovery for individuals with OUD \cite{meaningful_employment}. 

Although transportation barriers have been recognized as significant obstacles to general healthcare access in prior research \cite{transporation_barriers_general, traveling_towards_disease,one_in_five_travel,evaluating_barriers}, the role of household vehicle ownership in the opioid crisis remains largely underexplored. Lack of a household vehicle can lead to missed or delayed healthcare appointments, increased health expenditures, and poorer health outcomes \cite{aha}. These issues are particularly significant in the daily life of individuals struggling with OUD. In opioid-related emergencies, such as overdoses, direct access to transportation becomes vital, as the lack of it can delay life-saving care. Additionally, many underserved communities facing opioid challenges often lack adequate transportation, forcing individuals without vehicles to rely on limited local resources \cite{travel_barriers_disadvantaged, transporation_barriers_general}. For example, research has documented an alarming rise in opioid-related overdose deaths within Black urban populations, partly attributing this trend to the lack of available transportation in these communities \cite{urban_black_communities}.

In addition to highlighting the importance of unemployment and lack of vehicle access in opioid-related mortality, this study also underscores the detrimental impact of the COVID-19 pandemic on the national opioid crisis. While this impact has been previously discussed \cite{opioids_and_svi_and_covid,pnas_covid19,covid_transportation_barriers}, our work adds to the conversation by providing novel support. This is seen in the performance of our models, where we observe a notable increase in errors beginning in 2020 and persisting through 2022. This trend is significant because the autoencoder retains learned temporal patterns, whereas XGBoost treats each year of data independently without learning across time. The fact that both models exhibit similar error patterns during this period suggests that the opioid crisis worsened during the pandemic to such a degree that previous data patterns no longer provided useful predictive insights for the autoencoder.

The intensifying effects of the COVID-19 pandemic on the national opioid crisis were driven in part by two key factors: social isolation and unemployment \cite{covid_unemployment_loneliness,covid_unemployment_long_term_impact,social_isolation_covid_1, social_isolation_covid_2}. Unemployment in the United States surged to unprecedented levels during the pandemic, peaking at 14.8\% in April 2020 and gradually declining to 6.2\% in February 2021, still well above the pre-pandemic rate of 3.6\% in 2019 \cite{covid_unemployment_rates}. Additionally, the social isolation caused by pandemic lockdowns was shown to have detrimental effects on mental health and substance use outcomes \cite{social_isolation_covid_1, social_isolation_covid_2}. This social isolation, experienced widely during the pandemic, might reflect a similar form of isolation faced by individuals without access to reliable transportation, particularly in rural regions where public transit is limited and community resources are spread out. In this way, the COVID-19 pandemic highlights the destructive impact that both unemployment and lack of vehicle access can have on the national opioid crisis. 

Public health interventions could mitigate these impacts by prioritizing strategies that address both economic stability and transportation access. Workforce development programs that focus on providing stable, meaningful employment opportunities could play a critical role in reducing key stressors that contribute to opioid misuse, particularly in regions with high opioid-related mortality. Additionally, improving access to reliable and affordable transportation, especially in underserved communities and rural regions, could help alleviate the barriers faced by those without personal vehicles. By addressing these underlying structural challenges, targeted interventions could help reduce the impact of these two critical factors and may even promote long-term recovery in the nation’s most vulnerable areas.

This study has important limitations to note. Primarily, the missing mortality rates in the CDC WONDER dataset. Although we employed methods to address this, no approach can fully overcome the challenges posed by suppressed data. This limitation was sometimes evident when real data, missing in the previous year, became available the following year at a much higher rate than the previously imputed value. Such sharp increases occasionally led to large predictive errors in both models. Future research could apply the methodologies outlined in this study to a dataset with fewer missing values, if such data is available.

Another limitation is that the analysis is conducted at the county level. Using aggregate data may obscure important local differences, a phenomenon often referred to as the ecological fallacy. Additionally, this study does not stratify counties by urban or rural designation, but the same county-level characteristic can have markedly different implications in these two different contexts. For example, in urban areas such as New York City, low vehicle ownership rates may be a result of robust public transportation systems. Whereas in rural regions, such as Appalachia, lacking a vehicle can mean extreme isolation and limited access to healthcare. Future studies could explore such urban–rural dynamics to develop context-specific strategies for addressing transportation barriers.

Finally, while the thirteen SVI variables used in this study offer broad coverage of social factors, they do not capture all dimensions relevant to opioid-related mortality, such as: healthcare provider density, mental health service availability, incarceration rates, industry composition, long-term unemployment, college education rate, walkability, park access, or income inequality. For many of these factors, nationwide county-level data are either unavailable or limited to select years. These limitations render them impractical for the present design, as the timespan must align with that of the study period. Nevertheless, their omission constrains the model’s explanatory reach. If data availability permits, expanding the variable set in future work could help capture additional pathways and regional nuances in the opioid crisis.

\section{Conclusion}

Despite its limitations, this study provides valuable insight into some key structural factors driving the opioid crisis. By integrating exploratory data analysis with machine learning models, we identified vulnerabilities that may exacerbate opioid-related mortality when present at high levels, yet mitigate it when present at low levels. Across both methods of analysis, unemployment and lack of vehicle access consistently emerged as critical variables. These findings underscore the importance of addressing economic precarity and transportation barriers, particularly in regions most heavily affected by the epidemic. Additionally, while the limitations discussed above remain, the methods and results presented here offer a robust foundation for future research to build upon.

\subsection*{Acknowledgments}

\noindent The authors would like to thank Heidi A. Hanson and Andrew Farrell for their helpful comments and feedback throughout the entire study.

\subsection*{CRediT authorship contribution statement}

\noindent \textbf{Andrew Deas:} Writing – original draft, Writing – review and editing, Data curation, Methodology, Investigation, Formal analysis, Visualization, Software. \textbf{Adam Spannaus:} Writing – review and editing, Formal analysis, Methodology, Investigation, Supervision, Conceptualization, Funding acquisition. \textbf{Dakotah Maguire:} Writing – review and editing, Visualization, Methodology. \textbf{Jodie Trafton}: Writing – review and editing, Project administration, Funding acquisition. \textbf{Anuj J. Kapadia:} Writing – review and editing, Project administration, Funding acquisition, Supervision. \textbf{Vasileios Maroulas:} Writing – review and editing, Formal analysis, Methodology, Investigation, Supervision.

\subsection*{Data availability}

\noindent All datasets and code supporting this study are available in the `A-Deas/Opioid-Feature-Importance' Github repository: \url{https://github.com/A-Deas/Opioid-Feature-Importance}.

\subsection*{Competing interests}

\noindent The authors have no competing interests to declare.

\subsection*{Funding}

\noindent This quality improvement initiative is sponsored by the U.S. Department of Veterans Affairs (VA), is supported by and utilizes VA-funded resources from the Knowledge Discovery Infrastructure (KDI) at Oak Ridge National Laboratory and is backed by the Department of Energy (DOE) Office of Science. The manuscript has been authored by UT-Battelle LLC under contract DE-AC05-00OR22725 with the DOE. The US government retains a nonexclusive, paid-up, irrevocable, worldwide license to publish or reproduce this manuscript or allow others to do so for US government purposes. DOE will provide public access to these results in accordance with the DOE Public Access Plan (\url{http://energy.gov/downloads/doe-public-access-plan}). The initiative aimed is to improve care for Veterans within the Veterans Health Administration (VHA) and involved models specifically tailored to the VHA system. Therefore, it is not generalizable outside of VHA and is officially considered non-research by the VHA. Oak Ridge National Laboratory received Institutional Review Board (IRB) approval due to the secondary use of patient data, which complies with internal policies and standards.  The authors acknowledge the broader partnership and express gratitude to the Veterans receiving care at VA.

\def\UrlBreaks{\do\/\do-}
\bibliographystyle{unsrt}
\bibliography{references}

\end{document}